# Analytical solutions of the Dirac equation using the Tridiagonal Representation Approach: General study, limitations, and possible applications

Ibsal. A. Assi[a,†] and Hocine Bahlouli[a,b]

[a] *Physics Department, King Fahd University of Petroleum & Minerals, Dhahran 31261, Saudi Arabia*
[b] *Saudi Center for Theoretical Physics, P.O. Box 32741, Jeddah 21438, Saudi Arabia*

**Abstract**: This paper aims at extending our previous work on the solution of the one-dimensional Dirac equation using the Tridiagonal Representation Approach (TRA). In the approach, we expand the spinor wavefunction in terms of suitable square integrable basis functions that support a tridiagonal matrix representation of the wave operator. This will transform the problem from solving a system of coupled first order differential equations to solving an algebraic three-term recursion relation for the expansion coefficients of the wavefunction. In some cases, solutions to this recursion relation can be related to well-known classes of orthogonal polynomials whereas in other situations solutions represent new class of polynomials. In this work, we will discuss various solvable potentials that obey the tridiagonal representation requirement with special emphasis on simple cases with spin-symmetric and pseudospin-symmetric potential couplings. We conclude by mentioning some potential applications in graphene.

## 1. Introduction:

The basic equation of relativistic quantum mechanics was formulated by Paul Dirac in 1928 in a way consistent with special relativity [3]. This equation describes the behavior of weakly coupled electrons at high speeds or strongly coupled electrons such as in the case of core electron states in heavy atoms. Among the benefits of this relativistic formulation is the natural emergence of the electron spin and the prediction of the existence of an antiparticle partner to the electron, the positron, which was discovered experimentally few years later. The physics and mathematics of the Dirac equation is very rich, illuminating and provides a theoretical framework for different physical phenomena that are not present in the nonrelativistic regime such as the Klein paradox [4]. In addition, Dirac equation emerges in the study of the transport properties in graphene, which makes it important for future applications [5, 6].

Exact solutions of the Dirac equation with a given potential configuration are limited and not trivial compared to the nonrelativistic Schrödinger equation. In fact, the Dirac Hamiltonian being a matrix in the spinor space allows for more structure in the potential interaction. The terminology given to relativistic problems such as the "Dirac-Coulomb,"

† Corresponding Author: Email Address: g201403420@kfupm.edu.sa
Present Address: King Fahd University of Petroleum and Minerals, Dhahran 31261, Saudi Arabia.

"Dirac-Oscillator," "Dirac-Morse," …etc. refers to the Dirac equation that reduces to an effective Schrödinger-like equation with the named potential for the large spinor component. Different approaches were developed to generate exact solutions to the Dirac equation such as supersymmetric quantum mechanics and factorization method to mention only few [7-13].

This paper is an expanded version of our letters [1, 2] with further developments and applications in which we use the J-matrix inspired Tridiagonal Representation Approach (TRA). The basic idea of the approach is to write the spinor wavefunction as a bounded infinite series with respect to a suitably chosen square integrable basis function as $|\psi_\varepsilon(x)\rangle = \sum_m f_m(\varepsilon)|\phi_m(x)\rangle$, where $\{f_m(\varepsilon)\}_{m=0}^{\infty}$ is a set of expansion coefficients that are functions of the energy and potential parameters and $\{|\phi_m(x)\rangle\}_{m=0}^{\infty}$ is a complete set of spinor basis functions that carry only kinematic information. Using this form of the spinor wavefunction in the stationary wave equation, $(H-\varepsilon)|\psi\rangle = J|\psi\rangle = 0$, where $H$ is the Dirac Hamiltonian and requiring that the matrix representation of the wave operator, $J_{n,m} = \langle\phi_n|(H-\varepsilon)|\phi_m\rangle$, be tridiagonal and symmetric then this transforms the wave equation to the following three-term recursion relation for $\{f_m(\varepsilon)\}_{m=0}^{\infty}$ [14]:

$$J_{n,n}\ f_n(\varepsilon) + J_{n,n-1}\ f_{n-1}(\varepsilon) + J_{n,n+1}\ f_{n+1}(\varepsilon) = 0, \tag{1.1}$$

Thus, the problem now reduces to solving this three-term recursion relation which is equivalent to solving the original wave equation. Of course, this equation can be solved in different ways in mathematics [15, 32]. Sometimes solutions of Eq. (1.1) can be written in a closed form by direct comparison to well-known orthogonal polynomials. However, in other cases this recursion relation does not correspond to any of the known orthogonal polynomials giving rise to new class of orthogonal polynomials. The challenge will then be to write these solutions in a closed form and find the properties of the associated orthogonal polynomials such as the weight function, generating function, spectrum formula, asymptotics, zeros, etc.

In the following section, we introduce the general formulation of the problem and show how to calculate the matrix elements of the Dirac wave operator in a general basis. Then we consider two choices of bases depending on the configuration space of the problem. One is written in terms of the Jacobi polynomials and the other is written in terms of Laguerre polynomials. In the third section, we present various examples of exactly solvable potentials with special focus on possible applications. We give our conclusions and discuss future work in the last section.

## 2. Theoretical Modeling:

The most general form of 1D time-independent Dirac equation in the presence of scalar potential $S(x)$, two-component (time, space) vector potential $\vec{A}(x) = (V(x), U(x))$ and pseudo-scalar potential $W(x)$ can be written in the following form (in the relativistic units $\hbar = c = 1$):

$$\begin{pmatrix} m+S(x)+V(x) & -\frac{d}{dx}+W(x) \\ \frac{d}{dx}+W(x) & -m-S(x)+V(x) \end{pmatrix}\begin{pmatrix} \psi^{+}(x) \\ \psi^{-}(x) \end{pmatrix}=\varepsilon\begin{pmatrix} \psi^{+}(x) \\ \psi^{-}(x) \end{pmatrix}, \tag{2.1}$$

where we gauged away $U(x)$ to simplify the problem using a unitary transformation $\left|\psi^{\pm}(x)\right\rangle \rightarrow e^{-i\Lambda(x)}\left|\psi^{\pm}(x)\right\rangle$ provided that the common phase function $\Lambda(x)$ is chosen such that $d\Lambda/dx = U(x)$. We write $\psi^{\pm}(x)=\sum_n f_n(\varepsilon)\phi_n^{\pm}(x)$, where $\left\{\phi_n^{\pm}(x)\right\}_{n=0}^{\infty}$ is now a two-component spinor basis. The matrix elements, $J_{n,m}$, associated with (2.1) is written below:

$$J_{n,m}=\langle\phi_n|(H-\varepsilon)|\phi_m\rangle=\langle\phi_n^{+}|Q_{+}|\phi_m^{+}\rangle+\langle\phi_n^{-}|Q_{-}|\phi_m^{-}\rangle+\langle\phi_n^{+}|\left(-\tfrac{d}{dx}+W\right)|\phi_m^{-}\rangle+\langle\phi_n^{-}|\left(\tfrac{d}{dx}+W\right)|\phi_m^{+}\rangle, \tag{2.2}$$

where $Q_{\pm}=V(x)\pm S(x)\pm m-\varepsilon$.

In the presence of symmetry, we get more solutions to Eq. (2.1). These symmetries include the spin-symmetric coupling in which *V(x)=S(x)*, the pseudospin symmetric coupling which requires *V(x)= -S(x)*, and the presence of the scalar potential alone, i.e. *V(x)=W(x)=0*. We have discussed all the possible symmetries in section 3. In these symmetries, the problem reduces to solving an effective 1D Schrödinger-like equation which has been treated using the TRA in the past [14, 21, 22, 28]. In what follows, we will discuss the case in which there is no symmetry between the potentials in Eq. (2.1).

We start here by relating the spinor components using Eq. (2.1) as follows:

$$\psi^{-}(x)=\frac{1}{m+\varepsilon+S-V}\left[W+\frac{d}{dx}\right]\psi^{+}(x), \tag{2.3}$$

We refer sometimes to $\psi^{+}(x)$ as the "larger" component due to the dominator in (2.3). The general case we refer to holds when we have *V(x)* alone or *V(x)* and *S(x)* together with or without *W(x)* with no symmetry between *V(x)* and *S(x)*. To simplify the problem, we relate the corresponding basis functions through the *kinetic balance relation* [20]:

$$\phi_n^{-}(x)=\frac{\eta}{m+\varepsilon}\left[W+\frac{d}{dx}\right]\phi_n^{+}(x), \tag{2.4}$$

where $\eta$ is just a real dimensionless constant. Using Eq. (2.4) in (2.2) with the coordinate transformation $x\rightarrow y(x)$, we can write the matrix elements of the wave operator, J, as follows:

$$\begin{aligned} J_{n,m}=&-\frac{\eta}{\varepsilon+m}\langle\phi_n^{+}|y'^2\left[\frac{d^2}{dy^2}+\frac{y''}{y'^2}\frac{d}{dy}\right]|\phi_m^{+}\rangle \\ &+\frac{\eta}{\varepsilon+m}\langle\phi_n^{+}|\left[\frac{\varepsilon+m}{\eta}Q_{+}+\left(W^2-W'\right)\right]|\phi_m^{+}\rangle \end{aligned}, \tag{2.5}$$

where the prime over the variable stands for the derivative with respect to x, i.e. $W'=\frac{dW}{dx}$

The coordinate transformation is made such that we make the domain of the Hamiltonian compatible with the domain of the basis functions, see Table 1. The next step is to use basis functions that make (2.5) tridiagonal and symmetric. We usually use two important bases which are mentioned below:

(1) Laguerre Basis:

$$\phi_n^+(y) = A_n y^\alpha e^{-\beta y} L_n^\nu(y), \tag{2.6}$$

where $L_n^\nu(y)$ is the generalized Laguerre polynomial of order $n$ in $y$, $A_n = \sqrt{\frac{\Gamma(n+1)}{\Gamma(n+\nu+1)}}$ is just a normalization constant, and $\{\alpha,\beta,\nu\}$ are real parameters.

(2) Jacobi Basis:

$$\phi_n^+(y) = A_n (1-y)^\alpha (1+y)^\beta P_n^{(\mu,\nu)}(y), \tag{2.7}$$

where $A_n = \sqrt{\frac{(2n+\mu+\nu+1)}{2^{\mu+\nu+1}} \frac{\Gamma(n+1)\Gamma(n+\mu+\nu+1)}{\Gamma(n+\nu+1)\Gamma(n+\mu+1)}}$ is just a normalization constant, $y \in [-1, 1]$ and $P_n^{(\mu,\nu)}(y)$ is Jacobi polynomial of order n. The parameters $\{\alpha,\beta,\mu,\nu\}$, with $\mu,\nu > -1$, are real numbers.

An interesting task for the motivated reader is to use these bases to verify that Eq. (2.3) leads to Eq. (2.4), i.e. the tridiagonal representation of $J_{n,m}$ cannot be made unless we use the kinetic balance relation. In this section, we do the calculations in Laguerre basis, while the calculations in Jacobi basis can be found in Appendix B. This gives the following form of the matrix elements:

$$\begin{aligned} J_{n,m} = &-\frac{\eta A_n A_m \kappa^2}{\varepsilon+m}\left\langle L_n^\nu \middle| y^\nu e^{-y}\left[ y\frac{d^2}{dy^2} + (\nu+1-y)\frac{d}{dy}\right] \middle| L_m^\nu \right\rangle \\ &-\frac{\eta A_n A_m \kappa^2}{\varepsilon+m}\left\langle L_n^\nu \middle| y^\nu e^{-y} \frac{y}{y'^2} G(y) \middle| L_n^\nu \right\rangle \end{aligned}, \tag{2.8}$$

Where,

$$G(y) = \frac{y}{y'^2}\left( W^2 - W' + \frac{(\varepsilon+m)}{\eta} Q_+ \right) - \frac{\alpha(\alpha+a-1)}{y} + 2\alpha\beta + a\beta - b\alpha - \beta(\beta-b)y, \tag{2.9}$$

Using the properties of $L_n^\nu$ (A1), (A3), and (A4), we impose the following constraints:

(I) We use the coordinate transformation that satisfies $y' = \kappa y^a e^{by}$, for reals $\{a,b,\kappa\}$. This form is compatible with the Laguerre weight function in (2.6) and eases the measure transformation.

(II) The parameters are constrained to be $2\alpha + a = \nu + 1$ and $2\beta - b = 1$, to ensure the tridiagonal representation of the first matrix element in (2.8).

(III) $G(y) = \rho y + \sigma$, for reals $\rho$ and $\sigma$, to ensure the tridiagonal representation of the last matrix element in (2.8).

Thus, the tridiagonal form of $J_{n,m}$ becomes:

$$J_{n,m}=\frac{\eta\kappa^2}{\varepsilon+m}\Big\{\Big[\big(2n+\nu+1\big)\rho+\sigma+n\Big]\delta_{n,m} \\ -\rho\Big[\sqrt{n\big(n+\nu\big)}\delta_{n,m+1}+\sqrt{(n+1)\big(n+\nu+1\big)}\delta_{n,m-1}\Big]\Big\}, \tag{2.10}$$

The potentials that allow (2.10) to hold must be chosen such that:

$$\rho y+\sigma=\frac{y}{y'^2}\left(W^2-W'+\frac{(\varepsilon+m)}{\eta}Q_+\right) \\ -\frac{\alpha(\alpha+a-1)}{y}+2\alpha\beta+a\beta-b\alpha-\beta\big(\beta-b\big)y, \tag{2.11}$$

We discuss various examples of solvable potentials for different situations in section 3.

## 3. Results and Discussions:

This section is divided into three parts organized as follows. In section 3.1, we expose different results related to graphene. In sections 3.2, we discuss the set of possible solvable potentials in presence of spin symmetries. Then we move to section 3.3 to expose some results on the general case.

### 3.1: Scalar Potential:

In this situation we consider *W(x)=V(x)=0*. Applying the unitary transformation $e^{i\frac{\pi}{4}\sigma_y}$, where $\sigma_y$ is the 2*x*2 Pauli matrix $\begin{pmatrix}0 & -i\\ i & 0\end{pmatrix}$, on Dirac Hamiltonian in Eq. (2.1), this gives the following form of the wave equation:

$$\begin{pmatrix}0 & \frac{d}{dx}+m+S\\ -\frac{d}{dx}+m+S & 0\end{pmatrix}\begin{pmatrix}\chi^+(x)\\ \chi^-(x)\end{pmatrix}=\varepsilon\begin{pmatrix}\chi^+(x)\\ \chi^-(x)\end{pmatrix}, \tag{3.1.1}$$

where $|\chi\rangle=e^{i\frac{\pi}{4}\sigma_y}|\psi\rangle$. The reason behind this transformation is that Eq. (3.1.1) is equivalent to Dirac-Weyl equation for an electron in graphene moving under the influence of an external magnetic field acting perpendicular to the plane of the graphene sheet. To see the correspondence, we write down the Dirac-Weyl equation which reads [26, 27][‡]:

$$v_F\vec{\sigma}.\left(\vec{p}+\frac{e}{c}\vec{A}\right)\left|\Psi\big(x,y\big)\right\rangle=E\left|\Psi\big(x,y\big)\right\rangle, \tag{3.1.2}$$

where $v_F$ is the Fermi speed, $\vec{\sigma}=\big(\sigma_x,\sigma_y\big)^T$ is the vector Pauli matrices, $\vec{p}=-i\hbar\left(\frac{\partial}{\partial x},\frac{\partial}{\partial y}\right)$ is the 2D momentum operator, *e* is the electron's charge, *c* is the speed of light, $\vec{A}$ is the two-vector potential, and $E$ is the energy eigenvalue. Now, choosing the *z*-axis normal to the graphene sheet, then the magnetic field could be generated from the two-vector potential in the Landau gauge $\vec{A}=\big(0,A_y\big(x\big),0\big)$ as

‡ In the presence of electric fields, we just add the vector potential to the Hamiltonian in (3.1.2).

$$\vec{A}=\left(0,A_y\left(x\right),0\right),\vec{B}=\left(0,0,B(x)\right),B\left(x\right)=\frac{dA_y}{dx}, \tag{3.1.3}$$

This gauge suggests that the spinor is separable as $\left|\Psi\left(x,y\right)\right\rangle=e^{iky}\left|\psi\left(x\right)\right\rangle$ (translational symmetry). This will reduce Eq. (3.1.2) to Eq. (3.1.1) with the following maps:

$$m\to k\,,S\to\frac{e}{c\hbar}A_y,\ \varepsilon=E/\hbar v_F\text{, and }\left|\chi\right\rangle\to\left|\psi\right\rangle$$

If we define $F\left(x\right)=m+S\left(x\right)$, we can now relate the spinor wavefunction components in Eq. (3.1.1) as follows:

$$\chi^{\pm}(x)=\frac{1}{\varepsilon}\left[\pm\tfrac{d}{dx}+F(x)\right]\chi^{\mp}(x), \tag{3.1.4}$$

The constraint in Eq. (3.1.4) allows us to break Dirac equation into two effective Schrödinger equations for each spinor component which we write down in compact form as follows:

$$\left[\frac{d^2}{dx^2}-2\tilde{U}+2\tilde{E}\right]\chi^{\pm}=0\,, \tag{3.1.5}$$

where $\tilde{U}=\frac{F^2\pm F'}{2}$, and $\tilde{E}=\frac{\varepsilon^2}{2}$, with $F'=\frac{dF}{dx}$. Consequently, we just need to solve the effective Schrödinger equation (3.1.5) for any component and find the other spinor component using the relation in (3.1.4). However, we need to stress that each solution of (3.1.5) will cover part of the energy space complementary to the other one. Luckily, Schrödinger equation has been treated in the past, using the TRA, by different authors including Alhaidari and Bahlouli [14, 21, 22, 28]. We have tabulated few of the solvable potentials of Eq. (3.1.5), which were treated by the TRA, in Table 3. We should point out that the situation with *S=V=0* is mathematically similar to the previous case which results, again, in having two Schrödinger equations for each spinor component as in Eq. (3.1.5) with $\tilde{U}=\left(W^2\pm W'\right)/2$ and $\tilde{E}=\left(\varepsilon^2-m^2\right)/2$. Next, we will discuss different situations that will be useful for graphene system.

As a first example, we consider the following hyperbolic magnetic field barrier:

$$B\left(x\right)=\frac{B_0}{\cosh^2\left(\alpha x\right)}, \tag{3.1.6}$$

where $B_0$ and $\alpha$ are constants. This case corresponds to $m=k$ and $S=S_0\tanh\left(\alpha x\right)$, with $S_0=eB_0/\alpha c\hbar$. Comparison to Schrödinger equation, this situation is equivalent to the following potential:

$$\tilde{U}\left(x\right)=\frac{1}{2}\left[\frac{S_0\left(\alpha-S_0\right)}{\cosh^2\left(\alpha x\right)}+k^2+S_0{}^2+2kS_0\tanh\left(\alpha x\right)\right], \tag{3.1.7}$$

This potential is called the hyperbolic Rosen-Morse potential which was treated using the TRA in [14]. Using the energy spectrum of this potential (See Table 3), we write the energy spectrum of an electron in graphene in this hyperbolic barrier as follows:

$$\varepsilon_n^2/\hbar^2 v_F^2 = k^2 + S_0{}^2 - \alpha^2\left[\left(n+\frac{1}{2}-|\gamma|\right)^2 + \left(\frac{kS_0}{\alpha^2}\right)^2\left(n+\frac{1}{2}-|\gamma|\right)^{-2}\right], \qquad (3.1.8)$$

where $\gamma^2 = \frac{S_0(S_0-\alpha)}{\alpha^2} + \frac{1}{4}$ and $S_0(S_0-\alpha) > -\frac{\alpha^2}{4}$. This result agrees with the result obtained in [25]. The upper component of the spinor wavefunction is now written as [14]:

$$\chi_n^+(x,y) = A_n(1+\tanh\alpha x)^{\nu_n/2}(1-\tanh\alpha x)^{\mu_n/2} P_n^{(\mu_n,\nu_n)}(\tanh\alpha x)e^{iky}, \qquad (3.1.9)$$

where $\mu_n = \frac{1}{\alpha}\sqrt{-2(\vartheta_n - kS_0)}$, $\nu_n = \frac{1}{\alpha}\sqrt{-2(\vartheta_n - kS_0)}$, $\vartheta_n = \frac{\varepsilon_n^2}{2\hbar^2 v_F^2}$, and $A_n = \sqrt{\frac{\alpha(2n+\mu+\nu+1)}{2^{\mu+\nu+1}}\frac{\Gamma(n+1)\Gamma(n+\mu+\nu+1)}{\Gamma(n+\nu+1)\Gamma(n+\mu+1)}}$ is just a normalization constant. The lower spinor component can be easily calculated using Eq. (3.1.4).

Another interesting example we mention here is the case when the magnetic barrier takes the following exponentially decaying form:

$$B = B_0 e^{-\alpha x}, \qquad (3.1.10)$$

where $B_0$ and $\alpha$ are constants with $\alpha > 0$. This case corresponds to a scalar potential of the form $S = S_0 e^{-\alpha x}$, where $S_0 = -eB_0/c\hbar\alpha$. Now, using Eq. (3.1.5), we find that the Schrödinger potential reads:

$$\tilde{U}(x) = \frac{1}{2}\left[k^2 + S_0^2 \mathrm{e}^{-2x\alpha} + S_0(2k-\alpha)\mathrm{e}^{-x\alpha}\right], \qquad (3.1.11)$$

This is simply the 1D Morse oscillator potential, up to a constant, which was treated using the TRA for Schrödinger equation by Alhaidari in [14]. Using the results in [14], we write the energy eigenvalues for Dirac-Weyl equation as follows:

$$\varepsilon_n^2/\hbar^2 v_F^2 = k^2 - \alpha^2\left(\frac{\gamma}{\mu} + n + \frac{1}{2}\right)^2, \qquad (3.1.12)$$

where $\gamma = S_0(2k-\alpha)/\alpha^2$ and $\mu = |2S_0/\alpha|$. The upper component of the spinor wavefunction reads:

$$\chi_n^+(x,y,) = A_n e^{iky} e^{-(\alpha\nu x + \mu e^{-\alpha x})/2} L_n^\nu\left(\mu e^{-\alpha x}\right), \qquad (3.1.13)$$

where $\nu = \left|\frac{2\gamma}{\mu} + 2n + 1\right|$. Our results in this example agrees with previous findings [25].

One last example we mention in this section is what we call the Hulthén barrier in which the magnetic field takes the following form:

$$B(x)=\frac{B_0 e^{\alpha x}}{\left(e^{\alpha x}-1\right)^2}, \tag{3.1.14}$$

where $B_0$ and $\alpha>0$ are constants. Following the same procedure, this will be the situation when the scalar potential is $S(x)=\frac{S_0}{e^{\alpha x}-1}$. The associated supersymmetric potential (Schrödinger potential) now reads:

$$\tilde{U}(x)=\frac{1}{2}\left[\frac{S_0\left(S_0-\alpha\right)}{(-1+\mathrm{e}^{x\alpha})^2}+\frac{S_0\left(2k-\alpha\right)}{(-1+\mathrm{e}^{x\alpha})}+k^2\right], \tag{3.1.15}$$

The potential in (3.1.15) is the generalized Hulthén potential which was treated in the TRA in [14]. Based on the results obtained in [14], we write the energy eigenvalue of Dirac-Weyl equation for this situation as follows:

$$\varepsilon_n^2/\hbar^2 v_F^2=k^2-\frac{\alpha^2}{4}\left[n+\frac{\nu+1}{2}+\frac{2(\gamma-\omega)/\alpha^2}{n+\frac{\nu+1}{2}}\right]^2, \tag{3.1.16}$$

where $\omega=\frac{S_0\left(S_0-\alpha\right)}{2}=\frac{\alpha^2}{2}\left(\frac{\nu^2-1}{4}\right)$, $\gamma=\frac{S_0\left(2k-\alpha\right)}{2}$, and $\nu>-1$. Now, we write down the spinor component as follows:

$$\chi_n^+(x,y)=A_n 2^{\frac{\nu+\mu_n+2}{2}}\left(1-e^{-\alpha x}\right)^{(\nu+1)/2}\left(e^{-\alpha x}\right)^{(\mu_n+1)/2} e^{iky} P_n^{(\mu_n,\nu)}\left(1-2e^{-\alpha x}\right), \tag{3.1.17}$$

where $\mu_n>-1$ and satisfies $\left(\frac{\mu_n+1}{2}\right)^2=\frac{1}{4}\left[n+\frac{\nu+1}{2}+\frac{2(\gamma-\omega)/\alpha^2}{n+\frac{\nu+1}{2}}\right]^2$. We leave it to the interested reader to calculate the lower component using Eq. (3.1.4). Up to our knowledge, this situation was not treated in the past. Note that for very large values of *x* the barrier will behave similarly to the previously mentioned case. It is obvious that all of the previous systems has finitely many bound states as shown in the spectrum formulas. We have also solved other interesting situations in which the magnetic field is constant, singular $\frac{1}{x^2}$, and few other cases are summarized in Table 4.

### 3.2: Spin-Symmetry and Pseudo-Spin Symmetry:

Defining $\Sigma=V+S$ and $\Delta=V-S$, we refer to the spin-symmetric coupling the situation where we have $\Delta=0$, and the pseudospin symmetric coupling for which $\Sigma=0$.[§] These cases are very useful in nuclear physics [16-19]. For the case when $\Delta=0$, we use Eq. (2.1) to write an equation for the upper spinor component which reads:

$$\left[-\frac{1}{2}\frac{d^2}{dx^2}+U_{ss}\right]\psi^+=E\psi^+, \tag{3.2.1}$$

where $U_{ss}=\frac{W^2-W'}{2}+(\varepsilon+m)V$, $E=\frac{\varepsilon^2-m^2}{2}$, and $W'=\frac{dW}{dx}$. Thus, we need to solve Schrödinger equation (3.2.1) for $\psi^+$ and then use (2.1) to compute $\psi^-$. As discussed in previous section, we will rely on the obtained solutions of Schrödinger equation using the TRA, where we tabulated few of these solvable potentials in Table 3, to obtain solvable potentials in these cases. Similarly, one can follow the same procedure for $\Sigma=0$, which results in an effective Schrödinger equation for the spinor lower component which is shown below:

$$\left[-\frac{1}{2}\frac{d^2}{dx^2}+U_{ps}\right]\psi^-=E\psi^-, \tag{3.2.2}$$

where $2U_{ps}=W^2+W'+2(\varepsilon-m)V$, and $2E=\varepsilon^2-m^2$. In what follows, we expose different examples in which (3.2.1) and (3.2.2) are exactly solvable in the TRA in the presence of the potentials $U_{ss}$ and $U_{ps}$, respectively.

[§] Sometimes we use $\Delta=C_{ss}$ and $\Sigma=C_{ps}$, where $C_{ss}$ and $C_{ps}$ are constants for spin symmetry and pseudospin symmetry, respectively.

The first example of solvable potentials we would like to mention here is the case when $V(x)=-S(x)=V_0x^2$ with *W(x)=0*. The effective Schrödinger equation in this case reads:

$$\left[-\frac{1}{2}\frac{d^2}{dx^2}+(\varepsilon-m)V_0x^2\right]\psi^-=E\psi^-, \tag{3.2.3}$$

where $2E=\varepsilon^2-m^2$. This is equivalent to Schrödinger equation for the Harmonic oscillator with "frequency" $\omega=\sqrt{2(\varepsilon-m)V_0}$. The basis components are written in Laguerre basis as $\phi_n^-(x)=A_n(\lambda x)^{\nu+1/2}e^{-\lambda^2x^2/2}L_n^\nu(\lambda^2x^2)$, which was treated in the TRA, see [14], with $\nu=\pm 1/2$. Using the energy spectrum for the Harmonic oscillator obtained in the TRA, we write the bound states spectrum formula for this potential configuration in Dirac equation as follows:

$$\varepsilon_n^2-m^2=\pm 4\sqrt{2(\varepsilon_n-m)V_0}\begin{cases} n+1/4 & ,\nu=-1/2 \\ n+3/4 & ,\nu=+1/2 \end{cases}, \tag{3.2.4}$$

where $\varepsilon>0, V_0>0$ (or $\varepsilon<0, V_0<0$), and $n=0,1,2,...$. It is well known that Laguerre and Hermite polynomials are related when $\nu=\pm 1/2$, see [15, 32]. The spinor wavefunction lower component for this case reads

$$\psi_n^-(x)=A_n(\lambda x)^{\nu+1/2}e^{-\lambda^2x^2/2}L_n^\nu(\lambda^2x^2), \tag{3.2.5}$$

where $A_n=\sqrt{\lambda\Gamma(n+1)/\Gamma(n+\nu+1)}$. The spinor upper component can be easily evaluated using Eq. (2.1). For applications, this situation can be modeled for an electron in graphene moving under the influence of linear electric and magnetic fields by considering the following map in Dirac-Weyl equation: $m\to k$, $S\to\frac{e}{c\hbar}A_y$, $V\to V/\hbar v_F$

, $\varepsilon = E/\hbar v_F$ . Moreover, this case has also been studied in nuclear physics and our results match with what other authors obtained, see for example, [29-31]. However, one can follow a similar procedure for the same oscillator potentials for spin-symmetric couplings, i.e. for $V(x)=S(x)=V_0x^2$, with $W(x)$=0, and obtain the spectrum to be similar to (3.2.4) with $\varepsilon - m \to \varepsilon + m$ under the square root.

Another example we would like to mention here is the case when we have spin-symmetric coupling with $V(x)=V_0/\cosh^2(\alpha x)$, and $W=W_0\tanh(\alpha x)$. Using Eq. (3.2.1), we write $U_{ss}(x)$ as follows:

$$U_{ss}(x)=\frac{W_0^2}{2}+\frac{(\varepsilon+m)V_0-\alpha W_0/2-W_0^2/2}{\cosh^2(\alpha x)}, \qquad (3.2.6)$$

This is a special form of Rosen-Morse potential which was treated in the TRA in [14]. Using the spectrum formula for $U_{ss}(x)$, we calculate the bound states spectrum formula of Dirac particle in this potential configuration to be:

$$\varepsilon_n^2=m^2+W_0^2-\lambda^2\left(n+\frac{1}{2}-|D/\lambda|\right)^2, \qquad (3.2.7)$$

where $D^2=W_0^2+\alpha W_0-2(\varepsilon+m)V_0+\lambda^2/4$, and $\varepsilon+m<\left(W_0^2+\alpha W_0+\alpha^2/4\right)/2V_0$. The upper spinor component is written below [14]:

$$\psi_n^+(x)=A_n(1-\tanh\lambda x)^{\nu/2}(1+\tanh\lambda x)^{\mu/2}P_n^{(\mu,\nu)}(\tanh\lambda x), \qquad (3.2.8)$$

where $A_n=\sqrt{\frac{\lambda(2n+\mu+\nu+1)}{2^{\mu+\nu+1}}\frac{\Gamma(n+1)\Gamma(n+\mu+\nu+1)}{\Gamma(n+\nu+1)\Gamma(n+\mu+1)}}$ , and $\mu=\mu_n=\frac{1}{\lambda}\sqrt{m^2-\varepsilon_n^2}=\nu_n$.
To avoid complex parameters in Jacobi polynomials we require that $0<\kappa<2m$, where $\kappa=\left(W_0^2+\alpha W_0+\alpha^2/4\right)/2V_0$. Thus, the condition in (3.2.7) requires $\varepsilon<m$, which means that this system has finitely many bound states. We leave it to the interested reader to calculate $\psi^-(x)$ using (3.2.8) in (2.1).

One last case we would like to discuss in this section is when $S(x)=V(x)=V_0\cos(\kappa x)$, and $W(x)=0$. The potential function $U_{ss}(x)$ for this case is $U_{ss}=(\varepsilon+m)V_0\cos(\kappa x)$. The Schrödinger equation in the presence of sinusoidal potential was studied by Alhaidari and Bahlouli in [28]. The spinor basis component is written in terms of Jacobi basis. By comparison, we write the J-matrix elements associated with this case as follows:

$$\kappa^2 J_{n,m}=\left[\varepsilon^2-m^2-\kappa^2\left(n+\frac{\mu+\nu+1}{2}\right)^2\right]\delta_{n,m}-(\varepsilon+m)V_0\left[\delta_{n,m+1}+\delta_{n,m-1}\right], \qquad (3.2.9)$$

where $\mu^2=\nu^2=1/4$. Using (3.2.9) in (1.1), we write the three-term recursion relation as:

$$\left[\varepsilon^2 - m^2 - \kappa^2\left(n+\frac{\mu+\nu+1}{2}\right)^2\right] f_n = (\varepsilon+m)V_0\left[\, f_{n-1} + f_{n+1}\right], \qquad (3.2.10)$$

Based on (3.2.10), we have exact solution of the expansion coefficients $\{f_n\}_{n=0}^{\infty}$ that can be evaluated exactly at any order *n* with initial conditions usually taken to be $f_{-1}=0,\ f_0=1$. Unfortunately, exact solutions of (3.2.10) cannot be written in a closed form as the recursion relation cannot be compared to any well-known class of orthogonal polynomials contrary to what we had in the previous examples. In fact, the solutions are referred to new polynomials which have been called "*dipole polynomials*" and have been found in different physical problems like electron in the dipole field and non-central potential problems [21, 22]. Moreover, the eigenstates can be evaluated at any order and the energy eigenvalues can be computed numerically with high accuracy. As an illustration, we have tabulated the lowest ten energy eigenvalues in **Table 2**. The spinor upper component for this system is written as $\psi^+(x,\varepsilon_n)=\sum_j f_j(\varepsilon_n)\phi_j^+(x)$, where $\{f_j(\varepsilon_n)\}_{j=0}^{\infty}$ are solutions to (3.2.10) and $\phi_j^+(x)$ is written in terms of Jacobi basis below:

$$\phi_j^+(x) = A_j\left(1+\cos\kappa x\right)^{\frac{2\nu+1}{4}}\left(1-\cos\kappa x\right)^{\frac{2\mu+1}{4}} P_j^{(\mu,\nu)}\left(\cos\kappa x\right), \qquad (3.2.11)$$

where $A_n=\sqrt{\frac{\kappa(2n+\mu+\nu+1)}{2^{\mu+\nu+1}}\frac{\Gamma(n+1)\Gamma(n+\mu+\nu+1)}{\Gamma(n+\nu+1)\Gamma(n+\mu+1)}}$. Similarly, we can follow this procedure to find solutions for the pseudo-spin-symmetric coupling for the same sinusoidal potentials. For more solvable potentials in the presence of spin symmetries, we have tabulated more results in Table 5.

## 3.3. General Case:

An interesting example we mention here is when *S=W=0,* and $V(x)=V_0x^2$, which can be modeled for an electron in graphene moving under the influence of a parabolic electrostatic barrier. Using the transformation $y=\left(\frac{1}{2}\kappa x\right)^2$ in (2.11), we obtain $\rho=\frac{4(\varepsilon+m)V_0}{\eta\kappa^4}-\frac{1}{4}$, and $\sigma=\frac{m^2-\varepsilon^2}{\eta\kappa^2}+\frac{\nu+1}{2}$ for $\nu=\pm 1/2$. The J-matrix for this situation is given below:

$$J_{n,m}=\frac{\eta\kappa^2}{\varepsilon+m}\left\{\left[\left(n+\frac{\nu+1}{2}\right)\left(\frac{8(\varepsilon+m)V_0}{\eta\kappa^4}+\frac{1}{2}\right)+\frac{m^2-\varepsilon^2}{\eta\kappa^2}\right]\delta_{n,m} -\left(\frac{4(\varepsilon+m)V_0}{\eta\kappa^4}-\frac{1}{4}\right)\left[\sqrt{n(n+\nu)}\,\delta_{n,m+1}+\sqrt{(n+1)(n+\nu+1)}\,\delta_{n,m-1}\right]\right\}, \qquad (3.3.1)$$

Using Eq. (1.1), we write the three-term recursion relation as follows:

$$\frac{m^2-\varepsilon^2}{\eta\kappa^2}f_n=-\left(n+\frac{\nu+1}{2}\right)\left(\frac{8(\varepsilon+m)V_0}{\eta\kappa^4}+\frac{1}{2}\right)f_n$$
$$+\frac{1}{2}\left(\frac{8(\varepsilon+m)V_0}{\eta\kappa^4}-\frac{1}{2}\right)\left[\sqrt{n(n+\nu)}f_{n-1}+\sqrt{(n+1)(n+\nu+1)}f_{n+1}\right], \quad (3.3.2)$$

by comparison with the three-term recursion relation of Meixner-Pollaczek polynomials (A13), we find that the solutions to Eq. (3.3.2) are the normalized Meixner-Pollaczek polynomials. Using the infinite spectrum formula of these polynomials (A13), we obtain the following bound states spectrum for $\nu=1/2$:

$$\left[m^2-\varepsilon_n^2\right]=\pm 4\sqrt{(\varepsilon_n-m)\eta V_0}\left(n+\frac{3}{4}\right), \quad (3.3.3)$$

where $\varepsilon>0, V_0>0$ (or $\varepsilon<0, V_0<0$), and $\varepsilon>m+\eta\kappa^4/32V_0$. The upper spinor wavefunction associated with this case is given below:

$$\psi^+(x,\varepsilon_n)=\frac{\kappa}{2}xe^{-\kappa^2x^2/8}\sum_j A_j f_j(\varepsilon_n)L_j^\nu(\kappa^2x^2/4), \quad (3.3.4)$$

where $A_j=\sqrt{\Gamma(j+1)/\Gamma(j+3/2)}$, $f_n=\sqrt{\omega(\xi)}P_n^{\frac{3}{4}}(\xi;\theta)$, where $P_n^{\frac{3}{4}}(\xi;\theta)$ is the Meixner-Pollaczek polynomial of order n in $\xi$, $\omega(\xi)$ is its weight function (A11), with $\xi=\frac{m^2-\varepsilon^2}{4\sqrt{(\varepsilon+m)\eta V_0}}$ and $\cosh\theta=\left(\frac{8(\varepsilon+m)V_0}{\eta\kappa^4}+\frac{1}{2}\right)/\left(\frac{8(\varepsilon+m)V_0}{\eta\kappa^4}-\frac{1}{2}\right)$. The spinor wavefunction lower component can be easily obtained using the above result in Eq. (2.1).

We will not be able exhaust all solvable potentials in this section, but one can follow the same procedure to obtain different solvable potentials like $V(x)=V_0e^{-\kappa x}$, $V(x)=V_0\sin(\kappa x)$, and others. We should point out here that we have used the kinetic balance relation [20] in our calculations which allow analytical solutions of the wave equation in the none simple cases. This relation is based on nonrelativistic approximation which usually gives one energy solution either the "positive energy sector" or the "negative energy sector". The interested reader is advised to refer to [20] and references therein.

## 4. Conclusions and Future Recommendations:

We have solved the one dimensional Dirac equation using the Tridiagonal Representation Approach (TRA). This approach, even limited, provides a very easy and handy approach to find analytical solutions to a certain class of solvable potentials for the 1D Dirac equation. In the presence of symmetry between the potential components in Dirac equation, the problem can be reduced to solving an effective Schrodinger-like equation which was treated previously using the TRA [14, 21, 22, 28]. The solvable potential configurations we obtained have been discussed in details in section 3. As a potential application of our analytical results we have mentioned in section 3 that some of our results can be used directly in graphene to treat electrons subject to electrostatic or magneto static (or both) barriers, a subject of major importance in recent graphene literature. Finally, we would like to express our interest in extending our approach to Dirac equation in higher dimensions.

**Acknowledgements**:

The authors would like to thank King Fahd University of Petroleum and Minerals for their support under research group project RG1502-1 and RG1502-2. We also acknowledge the material support of the Saudi Center for Theoretical Physics (SCTP). Our deep appreciation goes to Prof. Abdulaziz Alhaidari under whose guidance this research has been performed.

## Appendix A: Orthogonal Polynomials

### A1. The Generalized Laguerre Polynomials:

Theses polynomials are solutions of the following second order differential equation:

$$\left\{ y\frac{d^2}{dy^2} + (\nu+1-y)\frac{d}{dy} + n \right\} L_n^\nu(y) = 0\,, \tag{A1}$$

where $y \in [0,\infty[$, $\nu > -1$, and $n \in Z^+$.

These polynomials satisfy the following useful properties:

$$y\frac{d}{dy}L_n^\nu(y) = nL_n^\nu(y) - (n+\nu)L_{n-1}^\nu(y)\,, \tag{A2}$$

$$yL_n^\nu(y) = (2n+\nu+1)L_n^\nu(y) - (n+\nu)L_{n-1}^\nu(y) - (n+1)L_{n+1}^\nu(y)\,, \tag{A3}$$

$$\int_0^\infty y^\nu e^{-y} L_n^\nu(y) L_m^\nu(y)dy = \frac{\Gamma(n+\nu+1)}{n!}\delta_{n,m}\,, \tag{A4}$$

### A2. The Jacobi Polynomials:

Jacobi polynomials $P_n^{(\mu,\nu)}(y)$ defined on $[-1, 1]$ are solutions to the following ODE:

$$\left\{ (1-y^2)\frac{d^2}{dy^2} - \left[(\mu+\nu+2)y + \mu - \nu\right]\frac{d}{dy} + n(n+\mu+\nu+1) \right\} P_n^{(\mu,\nu)}(y) = 0\,, \tag{A5}$$

Theses polynomials satisfy the following recursion properties:

$$(1-y^2)\frac{dP_n^{(\mu,\nu)}}{dy} = -n\left( y + \frac{\nu-\mu}{2n+\nu+\mu} \right) P_n^{(\mu,\nu)} + 2\frac{(n+\nu)(n+\mu)}{2n+\mu+\nu} P_n^{(\mu,\nu)}\,, \tag{A6}$$

$$y P_n^{(\mu,\nu)}(y) = \frac{\nu^2-\mu^2}{(2n+\mu+\nu)(2n+\mu+\nu+2)} P_n^{(\mu,\nu)}(y) + \frac{2(n+\nu)(n+\mu)}{(2n+\mu+\nu)(2n+\mu+\nu+1)} P_{n-1}^{(\mu,\nu)}(y) + \frac{2(n+1)(n+\mu+\nu+1)}{(2n+\mu+\nu+1)(2n+\mu+\nu+2)} P_{n+1}^{(\mu,\nu)}(y) , \quad (A7)$$

$$\int_{-1}^{1} (1-y)^{\mu} (1+y)^{\nu} P_n^{(\mu,\nu)} P_m^{(\mu,\nu)} dy = \frac{2^{\mu+\nu+1}}{2n+\mu+\nu+1} \frac{\Gamma(n+\mu+1)\Gamma(n+\nu+1)}{\Gamma(n+\mu+\nu+1)n!} \delta_{n,m} , \quad (A8)$$

**A3. The Meixner-Pollaczek Polynomials:**

These polynomials are defined in terms of the hypergeometric function as follows [23, 24]:

$$P_n^{\mu}(y;\theta) = \sqrt{\frac{\Gamma(n+2\mu)}{\Gamma(2\mu)\Gamma(n+1)}} e^{in\theta} {}_2F_1\left(-n, \mu+iy; 2\mu; 1-e^{-2i\theta}\right), \quad (A9)$$

where $y \in (-\infty, \infty)$, $\mu > 0$ and $0 < \theta < \pi$ .

These polynomials satisfy the following three-term recursion relation:

$$\left[z \sin(\theta)\right] P_n^{\mu}(z;\theta) = -\left[(n+\mu)\cos(\theta)\right] P_n^{\mu}(z;\theta) + \frac{1}{2}\left[\sqrt{n(n+2\mu-1)} P_{n-1}^{\mu}(z;\theta) + \sqrt{(n+1)(n+2\mu)} P_{n+1}^{\mu}(z;\theta)\right], \quad (A10)$$

The associated weight function reads as follows:

$$\omega(z) = \frac{1}{2\pi\Gamma(2\mu)} (2\sin\theta)^{2\mu} e^{(2\theta-\pi)z} |\Gamma(\mu+iz)|^2, \quad (A11)$$

To calculate the bound states we use the infinite spectrum formula associated with these polynomials which is given below:

$$z^2 = -(n+\mu)^2 , \quad (A12)$$

In some cases we obtain a recursion relation similar to (A10) but we cannot have the sine and the cosine between -1 and 1. In such circumstances we transform the problem by making a replacement θ→ iθ which makes (A10) reads:

$$\left[iz\sinh(\theta)\right]P_n^{\mu}(z;\theta)=-\left[(n+\mu)\cosh(\theta)\right]P_n^{\mu}(z;\theta)$$
$$+\frac{1}{2}\left[\sqrt{n(n+2\mu-1)}P_{n-1}^{\mu}(z;\theta)+\sqrt{(n+1)(n+2\mu)}P_{n+1}^{\mu}(z;\theta)\right], \quad \text{(A13)}$$

## Appendix B: The Jacobi Basis

Jacobi basis functions are defined in terms of Jacobi polynomials as follows:

$$\phi_n^{+}(y)=A_n(1-y)^{\alpha}(1+y)^{\beta}P_n^{(\mu,\nu)}(y), \quad \text{(B1)}$$

where $A_n=\sqrt{\frac{(2n+\mu+\nu+1)}{2^{\mu+\nu+1}}\frac{\Gamma(n+1)\Gamma(n+\mu+\nu+1)}{\Gamma(n+\nu+1)\Gamma(n+\mu+1)}}$ is just a normalization constant, $y\in[-1,1]$ and $P_n^{(\mu,\nu)}(y)$ is Jacobi polynomial of order n. The parameters $\{\alpha,\beta,\mu,\nu\}$ will be constrained in order to satisfy the tridiagonal requirements. Following the same procedure as in the case of Laguerre basis we write the J-matrix as:

$$J_{n,m}=\frac{\kappa^2}{\varepsilon+m}n(n+\mu+\nu+1)\delta_{n,m}+\frac{\kappa^2A_nA_m}{\varepsilon+m}\left\langle L_n^{\nu}\right|y^{\nu}e^{-y}G(y)\left|L_m^{\nu}\right\rangle, \quad \text{(B2)}$$

where,

$$G(y)=\frac{1-y^2}{y'^2}\left[W^2-W'+2(\varepsilon+m)V(x)+m^2-\varepsilon^2\right]$$
$$-\left[\beta(\beta+b-1)\frac{1-y}{1+y}+\alpha(\alpha+a-1)\frac{1+y}{1-y}-2\alpha\beta-b\alpha-a\beta\right], \quad \text{(B3)}$$

and we used the properties of $P_n^{(\mu,\nu)}(y)$ to impose the following constraints:

(I) $2\beta+b=\nu+1$ and $2\alpha+a=\mu+1$

(II) The coordinate transformation is chosen such that $\frac{dy}{dx}=\kappa(1-y)^{a}(1+y)^{b}$.[**]

Recall that our main objective is to make (B2) tridiagonal and symmetric which can be achieved, again, by the linearity of (B3), that is:

$$\rho y+\sigma=\frac{1-y^2}{y'^2}\left[W^2-W'+2(\varepsilon+m)V(x)+m^2-\varepsilon^2\right]$$
$$-\left[\beta(\beta+b-1)\frac{1-y}{1+y}+\alpha(\alpha+a-1)\frac{1+y}{1-y}-2\alpha\beta-b\alpha-a\beta\right], \quad \text{(B4)}$$

For reals $\rho$ and $\sigma$. The condition (B4) is tricky and can only be achieved for certain coordinate transformations, certain choice of parameters and potential configurations. Section 3 in this paper discusses few examples of solvable potentials that satisfy (B4).

[**] This choice of transformation where $y'$ is compatible with the weight function of $P_n^{(\mu,\nu)}$ ease the analytical calculations.

Once this linearity achieved, we can write the J-matrix in its tridiagonal symmetric form as follows:

$$J_{n,m} = \frac{\kappa^2}{\varepsilon + m}\left\{\left[n(n+\mu+\nu+1)+\sigma+\rho\frac{\nu^2-\mu^2}{(2n+\mu+\nu)(2n+\mu+\nu+2)}\right]\delta_{n,m}\right.$$
$$+\frac{2\rho}{2n+\mu+\nu}\sqrt{\frac{n(n+\mu)(n+\nu)(n+\mu+\nu)}{(2n+\mu+\nu-1)(2n+\mu+\nu+1)}}\delta_{n,m+1} \quad , \qquad \text{(B5)}$$
$$\left.+\frac{2\rho}{2n+\mu+\nu+2}\sqrt{\frac{(n+1)(n+\mu+1)(n+\nu+1)(n+\mu+\nu+1)}{(2n+\mu+\nu+1)(2n+\mu+\nu+3)}}\delta_{n,m-1}\right\}$$

**Table Captions:**

**Table 1:** Few interesting coordinate transformations that are very useful in our approach which will be used to obtain different class of solvable potentials.

**Table 2:** The first ten positive and negative energy solutions to Eq. (3.23). Here we took $m=1, V_0=0.5$, $\mu=\nu=1/2$, and $\kappa=1.5, 0.1$.

**Table 3:** Some of the solvable potentials for Schrödinger equation which were obtained in the past using the TRA [14, 22, 28].

**Table 4:** List of magnetic field configurations in graphene with the energy eigenvalue and the supersymmetric potential (Schrödinger potentials) of each case [14].

**Table 5:** Few examples of solvable potentials in the spin-symmetric coupling with the bound states spectrum formula and spinor wavefunction upper component for each [14]. One can obtain similar results in the pseudo-spin symmetry.

**Table 1**

| Transformation Equation | Domain Mapping: $x \to y$ | Basis |
|---|---|---|
| $y = \lambda x$ | $[0,\infty[ \to [0,\infty[$ | Laguerre |
| $y = (\lambda x)^2$ | $\Re \to [0,\infty[$ | |
| $y = \mu e^{-\lambda x}$ | $\Re \to [0,\infty[$ | |
| $y = \tanh(\lambda x)$ | $\Re \to [-1,1]$ | Jacobi |
| $y = 1 - 2e^{-\lambda x}$ | $[0,\infty[ \to [-1,1]$ | |
| $y = \cos(\lambda x)$ | $[0,\pi/2] \to [-1,1]$ | |

**Table 2**

| | $\kappa = 1.5$ | | $\kappa = 0.1$ | |
|---|---|---|---|---|
| n | $\varepsilon^+$ | $\varepsilon^-$ | $\varepsilon^+$ | $\varepsilon^-$ |
| 0 | 1.36094 | -2.36093 | 0.177463 | -1.177460 |
| 1 | 2.70386 | -3.70384 | 0.353700 | -1.353680 |
| 2 | 4.13751 | -5.13747 | 0.493753 | -1.493710 |
| 3 | 5.60360 | -6.60352 | 0.610999 | -1.610920 |
| 4 | 7.08308 | -8.08296 | 0.711619 | -1.711500 |
| 5 | 8.56935 | -9.56917 | 0.798889 | -1.798710 |
| 6 | 10.05950 | -11.05930 | 0.874517 | -1.874280 |
| 7 | 11.55220 | -12.55180 | 0.939061 | -1.938750 |
| 8 | 13.04640 | -14.04600 | 0.993513 | -1.993110 |
| 9 | 14.54190 | -15.54140 | 1.045450 | -2.044960 |

## Table 3

| $V(x)$ | Domain ($x$) | $E_n$ | $\psi(x,E_n)=\sum_m f_m(E_n)\phi_m(x)$ | Constraints |
|---|---|---|---|---|
| $\frac{A}{x}+\frac{B+l(l+1)/2}{x^2}$ | $[0,\infty[$ | $-\frac{1}{2}\left(\frac{A}{n+\nu/2+1}\right)^2$ | $\phi_n(x)=A_n x^{1+\nu/2}e^{-x/2}L_n^{\nu}(x)$ | $\nu=-1+2\sqrt{\left\|(l+1/2)^2+2B\right\|}$<br>$\nu>-1$ |
| $\frac{1}{2}\lambda^4x^2+\frac{B+l(l+1)/2}{x^2}$ | $]-\infty,\infty[$ | $\lambda^2(2n+\nu+2)$ | $\phi_n(x)=A_n x^{\frac{\nu}{2}+\frac{3}{4}}e^{-x/2}L_n^{\nu}(x)$ | $\nu=-1+2\sqrt{\left\|(l+1/2)^2+2B\right\|}$<br>$\nu>-1$ |
| $V(x)=\frac{\lambda^2}{2}\left(Ae^{-\lambda x}+\left(\frac{\mu}{2}\right)^2e^{-2\lambda x}\right)$ | $]-\infty,\infty[$ | $-\frac{\lambda^2}{2}\left(\frac{A}{\mu}+n+\frac{1}{2}\right)^2$ | $\psi_n(x)=A_n e^{-\lambda\left\|\frac{A}{\mu}+n+\frac{1}{2}\right\|x}e^{-\mu e^{-\lambda x}/2}L_n^{\left\|\frac{A}{\mu}+n+\frac{1}{2}\right\|}\left(\mu e^{-\lambda x}\right)$ | $\mu,\lambda>0$ |
| $\frac{C}{\left(e^{\lambda x}-1\right)^2}+\frac{A}{e^{\lambda x}-1}$ | $[0,\infty[$ | $E_n=-\frac{\lambda^2}{8}\left[n+\frac{\nu+1}{2}+\frac{2(A-C)/\lambda^2}{n+\frac{\nu+1}{2}}\right]^2$ | $\psi_n(x)=A_n(1-y)^{(\nu+1)/2}(1+y)^{(\mu+1)/2}P_n^{(\mu,\nu)}(y)$ | $C=\frac{\lambda^2\left(\nu^2-1\right)}{8}$<br>$\mu,\nu>-1$<br>$y=1-2e^{-\lambda x}$ |
| $V(x)=C\tanh(\lambda x)$<br>$+\frac{A}{\cosh^2(\lambda x)}$ | $]-\infty,\infty[$ | $E_n=-\frac{\lambda^2}{2}\left[\vartheta_n^2+\left(\frac{C}{\lambda^2}\right)^2\vartheta_n^{-2}\right]$ | $\psi_n(x)=A_n(1-y)^{\nu_n/2}(1+y)^{\mu_n/2}P_n^{(\mu_n,\nu_n)}(y)$<br>$y=\tanh(\lambda x),\ \lambda>0$ | $C=(\lambda\mu/2)^2-(\lambda\nu/2)^2$<br>$\vartheta_n=\left(n+\frac{1}{2}-\left\|D/\lambda\right\|\right)$<br>$E_n=-(\lambda\mu/2)^2-(\lambda\nu/2)^2$ |
| $V_0\cos(k\lambda x)$ | $[0,L]$ | See [28] | $\psi_n(x)\propto P_n^{(\pm 1/2,\pm 1/2)}\left[\cos(k\lambda x)\right]$ | $\lambda=\pi/L$<br>$k=0,1,2,3,\ldots$ |

**Table 4**

| $B(x)$ | $\tilde{U}(x)$ | $\varepsilon_n^2$ |
|---|---|---|
| $B_0$ | $\frac{1}{2}(\gamma x+k)^2+\frac{\gamma}{2}$<br>$\gamma = eB_0/c\hbar$ | $\hbar^2 v_F^2\left[2\gamma+\hbar\omega(2n+1)\right]$<br>$\gamma^2 = m\omega^2$ |
| $\frac{B_0}{x^2}$ | $\frac{k^2}{2}+\frac{\gamma(\gamma-1)}{2x^2}+\frac{k\gamma}{x}$<br>$\gamma = -eB_0/c\hbar$ | $\hbar^2 v_F^2 k^2\left[1-\left(\frac{\gamma}{n+\nu/2+1}\right)^2\right]$<br>$\nu = 2(\gamma-1) > -1$ |
| $B(x)=\frac{B_0}{\cos^2(\lambda x)}$ | $\frac{k^2-S_0^2}{2}+kS_0\tan(\lambda x)+\frac{S_0(S_0+\lambda)}{2}\sec^2(\lambda x)$<br>where $S_0 = eB_0/c\hbar\lambda$ | $\hbar^2 v_F^2\left[k^2-S_0{}^2+\lambda^2\left(n+\frac{1}{2}-\lvert D/\lambda\rvert\right)^2 -\lambda^2\left(\frac{kS_0}{\lambda^2}\right)^2\left(n+\frac{1}{2}-\lvert D/\lambda\rvert\right)^{-2}\right]$<br>where $D^2 = S_0(S_0+\lambda)+\lambda^2/4$ and<br>$S_0(S_0+\lambda) > -\lambda^2/4$ |
| $B(x)=\frac{B_0}{\sinh^2(\lambda x)}$ | $\frac{k^2-S_0^2}{2}-kS_0\coth(\lambda x)+\frac{S_0(S_0-\lambda)}{2\sinh^2(\lambda x)}$<br>where $S_0 = -eB_0/c\hbar\lambda$ | $\hbar^2 v_F^2\left[k^2-S_0{}^2-\lambda^2\left(n+\frac{1}{2}-\lvert D/\lambda\rvert\right)^2 -\lambda^2\left(\frac{kS_0}{\lambda^2}\right)^2\left(n+\frac{1}{2}-\lvert D/\lambda\rvert\right)^{-2}\right]$<br>where $D^2 = S_0(S_0-\lambda)+\lambda^2/4$ and<br>$S_0(S_0-\lambda) > -\lambda^2/4$ |

## Table 5

| $V(x)=S(x)$ | $W(x)$ | $\varepsilon_n$ | $\psi_n^+(x)$ |
|---|---|---|---|
| $\frac{\lambda^2}{2}\left[Ae^{-\lambda x}+B^2e^{-2\lambda x}\right]$ | **0** | $\varepsilon_n^2-m^2=-\lambda^2\left[\frac{A}{B}\sqrt{(\varepsilon_n+m)}+n+\frac{1}{2}\right]^2$ | $\psi_n^+(x)=A_n\,\mathrm{e}^{-\lambda\left\lvert\frac{A}{B}\sqrt{(\varepsilon_n+m)}+n+\frac{1}{2}\right\rvert-\sqrt{B}e^{-\lambda x}}\,L_n^{\left\lvert\frac{2A}{B}\sqrt{(\varepsilon_n+m)}+2n+1\right\rvert}\left(2B\sqrt{(\varepsilon+m)}e^{-\lambda x}\right)$<br>$\lvert\varepsilon\rvert<m\,,B>0\,,\;A_n=\sqrt{\lambda\,\Gamma(n+1)/\Gamma(n+\nu+1)}$ |
| $V_0e^{-\lambda x}$ | $W_0e^{-\lambda x}$ | $\varepsilon_n^2-m^2=-\lambda^2\left[\frac{2V_0}{W_0\lambda}(\varepsilon_n+m)+n+\frac{3}{2}\right]^2$ | $\psi_n^+(x)=A_n\,\mathrm{e}^{-\lambda\left\lvert\frac{2V_0}{W_0\lambda}(\varepsilon+m)+n+\frac{3}{2}\right\rvert-W_0e^{-\lambda x}/\lambda}\,L_n^{\left\lvert\frac{4V_0}{W_0\lambda}(\varepsilon+m)+2n+3\right\rvert}\left(2W_0e^{-\lambda x}/\lambda\right)$<br>$\lvert\varepsilon\rvert<m\,,W_0>0\,,\;A_n=\sqrt{\lambda\,\Gamma(n+1)/\Gamma(n+\nu+1)}$ |
| $V=V_0\tanh(\lambda x)$ | $W=W_0\tanh(\lambda x)$ | $\varepsilon_n^2-m^2=W_0^2-\lambda^2\left[\vartheta_n^2+V_0^2\left(\frac{\varepsilon_n+m}{\lambda^2}\right)^2\vartheta_n^{-2}\right]$<br>$\vartheta_n=\left(n+\frac{1}{2}-\lvert D/\lambda\rvert\right)$<br>$D^2=W_0{}^2+\alpha\,W_0+\lambda^2/4$<br>$W_0(W_0+\alpha)>-\lambda^2/4$ | $\psi_n^+(x)=A_n(1+\tanh\lambda x)^{\nu_n/2}(1-\tanh\lambda x)^{\mu_n/2}P_n^{(\mu_n,\nu_n)}(\tanh\lambda x)$<br>$\mu_n=\sqrt{-\left[\varepsilon_n^2-m^2-2V_0(\varepsilon_n+m)\right]}/\lambda$<br>$\nu_n=\sqrt{-\left[\varepsilon_n^2-m^2+2V_0(\varepsilon_n+m)\right]}/\lambda$<br>$\mu_n,\nu_n>-1$ |
| $\frac{C}{\left(e^{-\lambda x}-1\right)^2}+\frac{A}{e^{-\lambda x}-1}$ | 0 | $\varepsilon_n^2=m^2-\frac{\lambda^2}{4}\left[n+\frac{\nu+1}{2}+\frac{2(\varepsilon_n+m)(A-C)/\lambda^2}{n+\frac{\nu+1}{2}}\right]^2$<br>$\varepsilon_n^2-m^2=-\lambda^2(\mu+1)^2/4$ | $\psi_n(x)=A_n(1-y)^{(\nu+1)/2}(1+y)^{(\mu+1)/2}P_n^{(\mu,\nu)}(y)$<br>$y=1-2e^{-\lambda x}$<br>$\frac{\lambda^2(\nu^2-1)}{8}=C(\varepsilon_n+m)$ |